\documentclass[fleqn,usenatbib]{mnras}

\usepackage{newtxtext,newtxmath}

\usepackage[T1]{fontenc}

\usepackage{graphicx}	
\usepackage{amsmath}	
\usepackage{hyperref}
\usepackage{bm}
\usepackage[dvipsnames, table]{xcolor}
\usepackage{orcidlink}
\usepackage{float}
\usepackage{subcaption}

\newcommand{\bq}{\boldsymbol q}

\newcommand{\bk}{\boldsymbol k}
\newcommand{\bPsi}{\boldsymbol{\Psi}}

\newcommand{\kmax}{k_{\rm max}}

\newcommand{\hMpc}{\,h\text{Mpc}^{-1}}

\title[NN emulator of the 2PCF]{Neural Network-based model of galaxy power spectrum: Fast full-shape galaxy power spectrum analysis}

\author[S. Trusov et al. ]{Svyatoslav Trusov$^{1}$\thanks{E-mail: strusov@lpnhe.in2p3.fr},
Pauline Zarrouk$^{1}$,
Shaun Cole$^{2}$
\\
$^{1}$ Laboratoire de Physique Nucl\'{e}aire et de Hautes Energies (LPNHE), CNRS/IN2P3 \& Sorbonne Universit\'{e}, 4 place Jussieu, 75005 Paris, France \\
$^{2}$ Institute for Computational Cosmology, Department of Physics, Durham University, South Road, Durham DH1 3LE, UK.
}
\date{Accepted XXX. Received YYY; in original form ZZZ}

\pubyear{2025}

\begin{document}
\label{firstpage}
\pagerange{\pageref{firstpage}--\pageref{lastpage}}
\maketitle

\begin{abstract}
We present a Neural Network based emulator for the galaxy redshift-space power spectrum that enables several orders of magnitude acceleration in the galaxy clustering parameter inference, while preserving  3$\sigma$ accuracy better than 0.5\% up to $k_{\mathrm{max}}$=0.25$\hMpc$ within $\Lambda$CDM and around 0.5\% $w_0$-$w_a$CDM. Our surrogate model only emulates the galaxy bias-invariant terms of 1-loop perturbation theory predictions, these terms are then combined analytically with galaxy bias terms, counter-terms and stochastic terms in order to obtain the non-linear redshift space galaxy power spectrum. This allows us to avoid any galaxy bias prescription in the training of the emulator, which makes it more flexible. Moreover, we include the redshift $z \in [0,1.4]$ in the training which further avoids the need for re-training the emulator. We showcase the performance of the emulator in recovering the cosmological parameters of $\Lambda$CDM by analysing the suite of 25 AbacusSummit simulations that mimic the DESI Luminous Red Galaxies at $z=0.5$ and $z=0.8$, together as the Emission Line Galaxies at $z=0.8$. We obtain similar performance in all cases, demonstrating the reliability of the emulator for any galaxy sample at any redshift in $0 < z < 1.4$. We will make our emulator public at github repository. 
\end{abstract}




\section{Introduction}

Spectroscopic galaxy surveys yield detailed three-dimensional maps of the cosmic large-scale structures (LSS) by mapping the distribution of several millions of galaxies in the sky. Such maps are now a well-established cosmological probe to understand our universe’s content and its mysterious late-time expansion. Standard galaxy clustering analyses rely on summarising the rich but noisy 3D information with 2-point statistics: 2-point correlation function (2PCF) and its Fourier transform the power spectrum. We exploit two main features in the galaxy two-point statistics in order to constrain the expansion history of the universe and the growth of structures: Baryon Acoustic Oscillations \citep[BAO, ][]{2005MNRAS.362..505C,2005ApJ...633..560E} and Redshift Space Distortions \citep[RSD, ][]{1987MNRAS.227....1K}. 
The first feature is the imprint on the galaxy clustering left by perturbations in the baryon-photon plasma of the early universe that propagated as sound waves until decoupling. It led to a characteristic scale that corresponds to the position of the BAO peak or wiggles in the galaxy two-point statistics and that can be used to measure the expansion rate of the universe across time.
The second feature introduces anisotropies in the full-shape of galaxy clustering due to the line-of-sight (LOS) component of galaxy peculiar velocities when inferring distances from redshifts. The sensitivity of galaxy clustering to structure growth through RSD allows us to perform direct tests of gravity, and thus to test the validity of General Relativity (GR) at cosmological scales. \citep[e.g. ][]{2008Natur.451..541G}. The standard method for galaxy Full-Shape analysis consists in compressing the observed multipoles into a set of three parameters: two scaling parameters parallel and perpendicular to the line-of-sight $\alpha_{\parallel}$ and $\alpha_{\perp}$ also called the Alcock-Paczynski parameters \citep{1979Natur.281..358A}, and the amplitude $f\sigma_8$ where $f$ is the linear growth rate of structures and $\sigma_8$ is the amplitude of linear matter power spectrum at 8 $\hMpc$ scales \citep{1980lssu.book.....P}; while keeping the linear power spectrum fixed (template-based approach). The constraints on this set of compressed parameters are then interpreted in terms of cosmological parameters of a given model, such as the so-called $\Lambda$CDM model. The latest state-of-the art standard galaxy clustering analyses have reached ~3\% precision on the equation of state of dark energy and a ~10\% precision on the growth rate of structures at 3 effective redshifts in the range $0.4 < z < 0.7$ \citep{2021PhRvD.103h3533A}. Forthcoming spectroscopic surveys with exquisite statistical power, such as the Dark Energy Spectroscopic Instrument \citep[DESI, ][]{2016arXiv161100036D}, promise advances on the nature of dark energy and validity of GR at cosmological scales. By collecting the spectra of about 40 million extragalactic galaxies and quasars in $0 < z < 3.5$, DESI will increase the number of measurements of the growth rate over redshift by a factor of 3 and improve the precision on cosmological parameters by a factor of 2-10 depending on the redshift bin.

Thanks to improvements in computing facilities, it is also more and more possible to directly vary the underlying parameters of a cosmological model to fit the observed two-point statistics. This approach is called direct fitting or Full-Modelling and has received lot of attention recently as it enables tighter constraints on some cosmological parameters without including CMB priors with respect to the standard template approach \citep{2020JCAP...05..042I,2020JCAP...05..005D}.
These standard analyses (either the standard template-based or Full-Modelling approach) are usually limited to scales of galaxy separation where we can use an analytic model of the redshift space two-point statistics based on perturbation theory (PT) in the mildly non-linear regime. Recent developments proposed to complement PT predictions with additional nuisance parameters to account for the small-scale physics and ensure that the models are not sensitive to the associated galaxy formation processes that can impact at quasi-linear scales. This extension is referred to as Effective
Field Theory \citep[EFT, e.g.][]{2015JCAP...09..014V}. In the Full-Modelling approach, when performing parameter inference, the shape of the linear power spectrum changes at each step of the Markov Chain Monte Carlo (MCMC) sampler. It implies computing the linear power spectrum using a Boltzmann code at each step, in addition to the calculations of the PT corrections. Therefore, the Full-Modelling approach is computationally very expensive, which motivates the need to accelerate the evaluation time of the underlying theoretical model, especially in the context of the unprecedented amount of data that is coming from the new generation galaxy surveys.

Such fast likelihood evaluation can be achieved by the use of an emulator which can approximate the predictions of a given summary statistic for a given set of cosmological parameters in a much more efficient way while preserving the accuracy of the model. One can use emulators based on either a Taylor series expansion such as in \cite{2023JCAP...06..005M}, Gaussian processes \citep[e.g.][]{2019ApJ...884...29N,2020MNRAS.497.2213M} or machine-learning algorithms \citep[e.g.][]{2023MNRAS.523.3219C,2022JCAP...04..056D,2022MNRAS.511.1771S}.

In this paper, we present a Neural-Network (NN) emulator for the public state-of-the-art Lagrangian perturbation theory based model called \textsc{velocileptors} that also includes EFT terms \citep{velocileptors1,velocileptors2}. In Section~\ref{sec:theory}, we review the theoretical background of the model in order to highlight the key quantities we want to emulate. In Section~\ref{sec:emulator}, we describe the NN based emulator and its performance in reproducing the reference non-linear power spectrum. In Section~\ref{sec:inference}, we present the simulations, methodology and results we obtain when performing a cosmological inference from Full-Modelling using either our NN-based emulator or the reference analytic model. To assess the performance of the emulator in constraining the parameters of LCDM, we use N-body simulations that reproduce the Luminous Red Galaxies (LRG) of DESI at redshifts $z=0.5$ and  $z=0.8$ and the Emission Line Galaxies (ELG) of DESI at redshift $z=0.8$. We conclude in Section~\ref{sec:concl}.


\section{From density contrasts to galaxy clustering}
\label{sec:theory}

In this section, we briefly describe the Lagrangian perturbation theory (LPT) we use as the theoretical model we choose to emulate. We also present the theory module of the analysis pipeline and the portion the emulator replaces to describe the main quantities we need to emulate for our chosen LPT. This is done to reduce the dimensionality of the input array and to avoid emulating dependencies on parameters that are not directly related to cosmology, such as galaxy biases and counter-terms. We will mostly follow the description from \cite{velocileptors1} and \cite{Matsubaru}.


\subsection{Lagrangian Perturbation Theory}

LPT tracks the trajectories $\boldsymbol{x}(\boldsymbol{q},t) = \boldsymbol{q}+\boldsymbol{\Psi}(\boldsymbol{q},t)$, of infinitesimal fluid elements originating at Lagrangian positions $\boldsymbol{q}$. This can be connected to the density contrast $\delta(\boldsymbol{x})$ in configuration and Fourier spaces as:
\begin{equation}
    1+\delta(\boldsymbol{x})=\int d^3\boldsymbol{q}\ \delta_D(\boldsymbol{x}-\boldsymbol{q}-\boldsymbol{\Psi}(\boldsymbol{q}))
\end{equation}
\begin{equation}
    (2\pi)^3 \delta_D(\boldsymbol{k})+\delta(\boldsymbol{k})=\int d^3\boldsymbol{q}\ e^{-i\boldsymbol{k}\cdot(\boldsymbol{q}+\boldsymbol{\Psi}(\boldsymbol{q}))}
\end{equation}
where $\delta_D(x)$ is the Dirac delta-function.

The equation of motion governing the evolution of $\boldsymbol{\Psi}$ under the influence of gravity can be written as:

\begin{equation}
    \ddot{\boldsymbol{\Psi}} +\mathcal{H}\dot{\boldsymbol{\Psi}}=-\nabla\Phi(\boldsymbol{x})
\end{equation}
where $\Phi(x)$ is the gravitational potential, dots represent derivatives with respect to the conformal time and $\mathcal{H} = aH$ is the conformal Hubble parameter. We adopt the approach of perturbation theory around the linear initial density contrast $\delta_0$ and solve this equation order-by-order by expanding $\boldsymbol{\Psi} = \boldsymbol{\Psi}^{(1)}+\boldsymbol{\Psi}^{(2)}+...$, where we can describe the terms in the expansion as:

\begin{multline}
    \boldsymbol{\Psi}^{(n)}_i(\boldsymbol{q}) = \frac{i^n}{n!} \int_{\boldsymbol{k},\boldsymbol{p}_1...\boldsymbol{p}_n} d^3\boldsymbol{k} \prod^n_{i=1} d^3\boldsymbol{p_i} e^{i\boldsymbol{k}\cdot\boldsymbol{q}} \times \\ \times \delta^D_{\boldsymbol{k}-\boldsymbol{p}}L^{(n)}_i(\boldsymbol{p}_1...\boldsymbol{p}_n)\Tilde{\delta}_0(\boldsymbol{p}_0)...\Tilde{\delta}_0(\boldsymbol{p}_n)
\end{multline}
where $L_i^{(n)}$ are the perturbation theory kernels, which are described in more detail in \cite{Matsubaru}, for instance. This also allows us to define the real-space pairwise displacement field as $\Delta_i = \Psi_{i}(\boldsymbol{q}_1) - \Psi_{i}(\boldsymbol{q}_2)$, which will be used later.

Cosmological surveys observe discrete tracers such as galaxies rather than the underlying matter distribution, therefore one needs to connect the statistical properties of galaxies with those of the matter density field. This connection, also called the galaxy bias model, encodes information about non-perturbative effects and baryonic physics. In the Lagrangian framework, we include a bias functional in the initial conditions, $F[\delta_0(\bq)]$, that relates the tracer overdensity field to the linear matter field in the form of a Taylor series. In Fourier space, this results in  
\begin{multline}
   (2\pi)^3\delta_D(\bk)+\delta_g(\bk) = \int d^3\bq\ F[\delta_0(\bq)]e^{-i\bk \cdot(\bq + \bPsi(\bq))} \nonumber \\
   F[\delta_0(\bq)] = 1 +  b_1\delta_0 + \frac{1}{2}b_2(\delta_0(\bq)^2 - \left\langle\delta_0^2\right\rangle) + b_s(s_0^2(\bq) - \left\langle s^2\right\rangle) + \\ + b_3 \mathcal{O}_3(\bq)
\end{multline}
where $b_1$ can be connected to the linear Eulerian bias $b_{1,\textrm{e}}$ as $b_1=b_{1,\textrm{e}}-1$, $s_0= (\partial_i\partial_j/\partial^2 - \delta_{ij}/3)\delta_0$ is the initial shear tensor, where we follow the notation of \cite{velocileptors1}. There is only one non-degenerate cubic contribution in the bias model at 1-loop order which we include schematically as $\mathcal{O}_3$. In this paper, we set $b_3 = 0$ and we refer to Maus et al. (in prep) for further tests of this assumption.





Galaxy redshifts contain two dominant contributions for galaxy clustering analysis, one that corresponds to the Hubble flow and another one that corresponds to the line-of-sight component of the galaxy peculiar velocity. The second contribution is responsible for the so-called redshift space distortions \citep[RSD][]{Kaiser} of the galaxy 2-point statistics. This effect needs to be accounted for in the model by boosting the displacement field $\bPsi$ along the line of sight $\hat{z}$ as follows: 
\begin{align}
    \bPsi_s = \bPsi + \frac{\hat{z}(\textbf{v}\cdot\hat{z})}{\mathcal{H}},
\end{align}
where $\textbf{v}$ is the galaxy peculiar velocity. In a matter-dominated universe, one can relate the displacement shift due to peculiar velocity to the linear growth rate of structures, $f$ such that for each of the perturbative kernels of order $n$ \citep{Matsubaru}:
\begin{align}
    \bPsi_s^{(n)} = \bPsi^{(n)} + nf(\hat{z}\cdot\bPsi^{(n)})\hat{z},
\end{align}
Eventually, we define the pairwise displacement field in redshift space as $\Delta_s = \Psi_s(\bq_1) - \Psi_s(\bq_2)$, and we can obtain the redshift-space galaxy power spectrum \cite{velocileptors1} as:
\begin{align}
    P_{g,s}(\bk) = \int d^3\bq \left\langle e^{i\bk \cdot (\bq + \Delta_s)}F(\bq_1)F(\bq_2) \right\rangle_{\bq = \bq_1-\bq_2}.
    \label{eq: VPint} 
\end{align}
where $F(\boldsymbol{q}) = F[\delta_0(\boldsymbol{q})]$.



In this paper, we consider the public state-of-the-art Effective Field Theory (EFT) code named \texttt{velocileptors}~\footnote{\url{https://github.com/sfschen/velocileptors}}~\citep{velocileptors1, velocileptors2} as our reference theory. This model is one of the EFT models used in DESI for the Full-Shape analysis of the DR1 galaxy samples. More precisely, we focus on the moment expansion model as implemented in the \texttt{MomentExpansion} module of \texttt{velocileptors}. 


\subsection{Moment expansion}
\label{sec:ME}

The redshift power spectrum can be expanded as:
\begin{equation}
    P_s(\boldsymbol{k}) = \sum^{\infty}_{n=0}\frac{i^n}{n!}k_{i_1}...k_{i_n}\Xi^{(n)}_{i_1...i_n}(\boldsymbol{k})
\end{equation}
where each of the moments $\Xi^{(n)}_{i_1...i_n}$ can be presented with density weighting, following \cite{velocileptors1} as:
\begin{equation}
    \Xi^{(n)}_{i_1...i_n} = \int d^3\boldsymbol{r} e^{i\boldsymbol{k}\cdot \boldsymbol{r}}\langle(1+\delta_1)(1+\delta_2)\Delta v_{i_1} ... \Delta v_{i_n}\rangle
\end{equation}
We can immediately notice that the zeroth moment is just a real-space power spectrum, while the first and second are the mean pairwise velocity $v_{12,i}$ and the velocity dispersion $\sigma_{12,ij}$ which are the main ingredients of the Moment Expansion model. 
We will not present the complete derivation, which can be found in \cite{velocileptors1, velocileptors2} but instead we will provide the final expressions only. Before that, we will define a shorthand notation for different correlators:
\begin{align*}
    U^{mn}_i &= \left\langle \delta^m(\boldsymbol{q}_1)\delta^n(\boldsymbol{q}_1)\Delta_i\right\rangle \\ A^{mn}_{ij} &= \left\langle  \delta^m(\boldsymbol{q}_1)\delta^n(\boldsymbol{q}_1)\Delta_i \Delta_j\right\rangle  \\ W_{ijk} &= \left\langle  \Delta_i \Delta_j\Delta_k\right\rangle
\end{align*}
Some of the correlators in the final expression will have linear and loop corrections separated, we will denote them that by a corresponding superscript.

We can expand the real-space galaxy power spectrum $P_g(k)$ into 12 cosmology-dependent terms multiplied by the biases $b_1, b_2, b_s, b_3$:
\begin{multline}
    P_g(k) = \int d^3\boldsymbol{q}e^{i\boldsymbol{k}\boldsymbol{q}}e^{-\frac{1}{2}k_ik_jA^{lin}_{ij}}\left\{1-\frac{1}{2}k_ik_jA^{\rm loop}{ij} + \frac{i}{6}k_ik_jk_kW_{ijk}\right. \\
    + b_1\left(2ik_iU_i-k_ik_jA^{10}_{ij}\right) + b_1^2\left(\xi_{\rm lin} +ik_iU^{11}_i-k_ik_jU^{\rm lin}_iU^{\rm lin}_j\right)+ \\
    + \frac{1}{2}b_2^2\xi^2_{\rm lin} + 2ib_1b_2\xi_{\rm lin}k_iU^{\rm lin}_i - b_2\left(k_ik_jU^{\rm lin}_i U^{\rm lin}_j + ik_iU^{20}_i  \right) + \\ + b_s \left(-k_ik_j\Upsilon_{ij} +2ik_iV^{10}_i \right) + 2ib_1b_sk_iV_i^{12} + b_2b_s\chi + b^2_s\zeta + \\ \left.+ 2ib_3 k_iU_{b_3,i}+2b_1b_3\theta + \alpha_Pk^2+...\right\} + R_h^3 = \\ = C_0 + b_1C_1 + b_1^2 C_2 + b_2^2 C_3 +b_1b_2 C_4 + b_2 C_5 + b_s C_6 + \\ + b_1b_s C_7 +  b_2b_s C_8 + b_s^2 C_9 + b_3 C_{10} + b_1b_3 C_{11} +\alpha_P k^2 + ...
\end{multline}

The mean pairwise velocity $v_{12}$ can be expanded in a similar fashion into 8 terms:
\begin{multline}
    \hat{n}_i v_{12,i}(\boldsymbol{k}) = \hat{n}_i\int d^3\boldsymbol{q}e^{i\boldsymbol{k}\boldsymbol{q}}e^{-\frac{1}{2}k_ik_jA^{lin}_{ij}}\left\{ik_j \dot{A}_{ji}-\frac{1}{2}k_jk_k\dot{W}_{jki} \right. + \\ + 2b_1\left(\dot{U}_i-k_kU_k^{\rm lin}k_j\dot{A}^{\rm lin}_{ji} + k_j\dot{A}^{10}_{ji}\right) + \\ + b_1^2\left( 2ik_jU^{\rm lin}_j\dot{U}^{\rm lin}_i + i\xi_{\rm lin}k_j\dot{A}^{\rm lin}_{ji}+\dot{U}^{11}  \right) + \\ + b_2\left(\dot{U}^{20}+2ik_jU^{\rm lin}_j \right) + 2b_1b_2\xi_{\rm lin}\dot{U}^{\rm lin}_i + 2b_s\left(\dot{V}^{10}_i + ik_j\dot{\Upsilon}_{ji}\right) + \\ \left. + 2b_1b_s\dot{V}^{12}_i + 2b_3\dot{U}_{b_3,i} + \alpha_vk_i+ \right\}... + R_h^4\sigma_v \\ = C_{12} + b_1 C_{13} + b_1^2 C_{14} + b_2 C_{15} + b_1 b_2 C_{16} + b_s C_{17} + b_1b_s C_{18} + \\ + b_3C_{19} + \alpha_{v}k_i + ...
\end{multline}

The pairwise velocity dispersion $\sigma_{12}$ in the same manner can be shown to be decomposed into 5 cosmology-dependent terms:
\begin{multline}
    \sigma_{12,ij}(\boldsymbol{k}) =  \int d^3\boldsymbol{q}e^{i\boldsymbol{k}\boldsymbol{q}}e^{-\frac{1}{2}k_ik_jA^{lin}_{ij}}\left\{ \ddot{A}_{ij} + ik_n\ddot{W}_{nij} - k_nk_m\dot{A}^{\rm lin}_{ni}\dot{A}^{\rm lin}_{mj}  \right.+\\+ b_1\left(2ik_nU_n^{\rm lin}\ddot{A}^{\rm lin}_{ij} + 2ik_n \left[\dot{A}^{\rm lin}_{ni}\dot{U}^{\rm lin}_j + \dot{A}^{\rm lin}_{nj}\dot{U}^{\rm lin}_i\right] + 2\ddot{A}^{10}_{ij}\right) + \\ + b1^2\left(\xi_{\rm lin}\ddot{A}^{\rm lin}_{ij} + 2 \dot{U}^{\rm lin}_i\dot{U}^{\rm lin}_j + \right) + 2 b_s \ddot{\Upsilon}_{ij} +\\ \left.+ \alpha_{\sigma}\delta_{ij} + \beta_{\sigma}\xi^2_{0,L}\left(\hat{q}_i\hat{q}_j -\frac{1}{3}\delta_{ij}\right) + ...\right\} + R^3_hs^2_v\delta_{ij}
\end{multline}
 The expression can be decomposed into transversal and longitudinal components as:
\begin{equation}
    \sigma_{ij} = \sigma_0(k)\delta_{ij} + \frac{3}{2}\sigma_2(k)\left(\hat{k}_i\hat{k}_j-\frac{1}{3}\delta_{ij}\right)
\end{equation}

Finally, adding the counter-terms $\alpha_i$, stochastic terms and the FOG effect $\sigma^2_{v}$, the total galaxy redshift-space power spectrum obtained using the Moment Expansion $P^{\rm ME}_{g,s}(\boldsymbol{k})$ is given by:
\begin{equation}
    \begin{split}
        P^{\rm ME}_{g,s}(\boldsymbol{k}) = \left( P(k) + i(k\mu)v_{12,\hat{n}}(\boldsymbol{k}) -\frac{(k\mu)^2}{2}\sigma^2_{12,\hat{n}\hat{n}}(\boldsymbol{k}) \right) + \\ +\left(\alpha_0 + \alpha_2\mu^2+\alpha_4\mu^4+...\right)k^2P_{\rm lin, Zel}(k)+R^3_h(1+\sigma^2_{v}(k\mu)^2+...)
    \end{split}
    \label{eq:ME-Pk}
\end{equation}

\subsection{Theory module of the analysis pipeline}

\begin{figure*}
    \centering
    \includegraphics[width = 1. \linewidth]{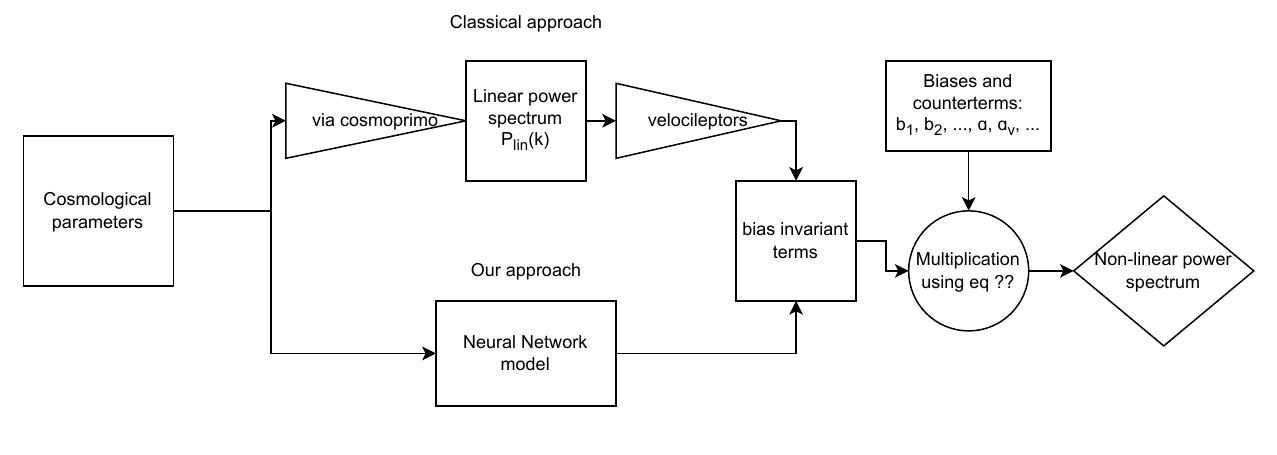}
    \caption{Schematic of the theory module in the analysis pipeline: the classical approach consists in first predicting the linear power spectrum and then computing the non-linear power spectrum with an EFT model such as \texttt{velocileptors}. We propose to replace the computation of the linear power spectrum and of the bias-invariant terms in the PT model by a neural-network emulator. These bias-invariant terms that depend only on the cosmological model are combined with the a set of bias and nuisance parameters, common to the \texttt{velocileptors} and NN pipeline,
    to predict the non-linear redshift-space galaxy power spectrum.}
    \label{fig:theory_module}
\end{figure*}

Fig.~\ref{fig:theory_module} presents the theory module of the analysis pipeline, or in other words how to predict the galaxy non-linear power spectrum multipoles from the input parameters which are the $\Lambda$CDM cosmological parameters and the redshift.
The upper branch corresponds to the default pipeline where we first compute the linear power spectrum using \texttt{cosmoprimo}\footnote{\url{https://github.com/cosmodesi/cosmoprimo}}, a python wrapper for \texttt{class}\footnote{\url{https://lesgourg.github.io/class_public/class.html}}, a simulator of the evolution of the linear cosmological perturbations, which is used as input for \texttt{velocileptors}. As shown in the previous section, we can split the contributions of each ingredient of the model between bias-invariant terms that depend on the cosmological model only and cosmology-independent terms that correspond to the bias model, counterterms, FOG and stochastic terms. Therefore, we can replace the generation of the linear power spectrum and the computation of the cosmoloy terms of the model by a neural network. This corresponds to the lower branch of Fig.~\ref{fig:theory_module}. Once the bias-independent terms are obtained with the neural network emulator, they can be added to the other terms in order to produce the non-linear power spectrum using Equation~\ref{eq:ME-Pk}.

By only emulating the 31 bias-independent terms which depend just on the cosmological parameters, our task is significantly simplified. Also emulating these terms avoids
the computational expensive bottleneck of computing the
 Fast Fourier Transforms required in \texttt{velocileptors}.
In our approach, the training sample for the emulation is based on \texttt{velocileptors} predictions but measurements from N-body simulations could be used instead as, for instance, in \cite{2020MNRAS.492.5754M}. Both approaches are limited to quasi-linear scales 
as the bias expansion is valid only on scales where the baryonic effects (such as active galactic nuclei feedback and ionising radiation) are small enough that they can be treated as perturbative corrections to the total power spectrum \citep[e.g. ][]{2018PhRvD..97f3526L,2019OJAp....2E...4C}. Moreover, in our approach, as the NN model is trained on PT predictions, it has the same range of validity as PT. We recall that we aim at significantly accelerating the cosmological inference and not at extracting information from the non-linear regime.

In the realistic surveys, however, the distances are usually converted from redshifts using a fiducial cosmology, which can be different from the underlying one. In order to account for this difference, we use the methodology presented in \cite{dAmico2020} .


\section{From analytical computations to neural networks}
\label{sec:emulator}

In this section, first, we describe the architecture of the neural network based emulator and then present its performance.

\subsection{Architecture of the emulator}

A fully connected neural network can be used to approximate a function $f$ such that $\mathbf{y} = f(\mathbf{x}|\boldsymbol{\theta})$,
where $\mathbf{x}$ represents the features of the data set, $\mathbf{y}$ the desired outputs, and $\boldsymbol{\theta}$ the free parameters of the network  which also can be referred to as trainable parameters. The optimal function $f$ is defined by the set of parameter values $\boldsymbol{\theta}$ that minimises the loss function (the form of which is discussed below). The loss function provides a measure of the performance of the model when evaluated on the data set.

\begin{figure}
    \centering
    \includegraphics[width = 1. \linewidth]{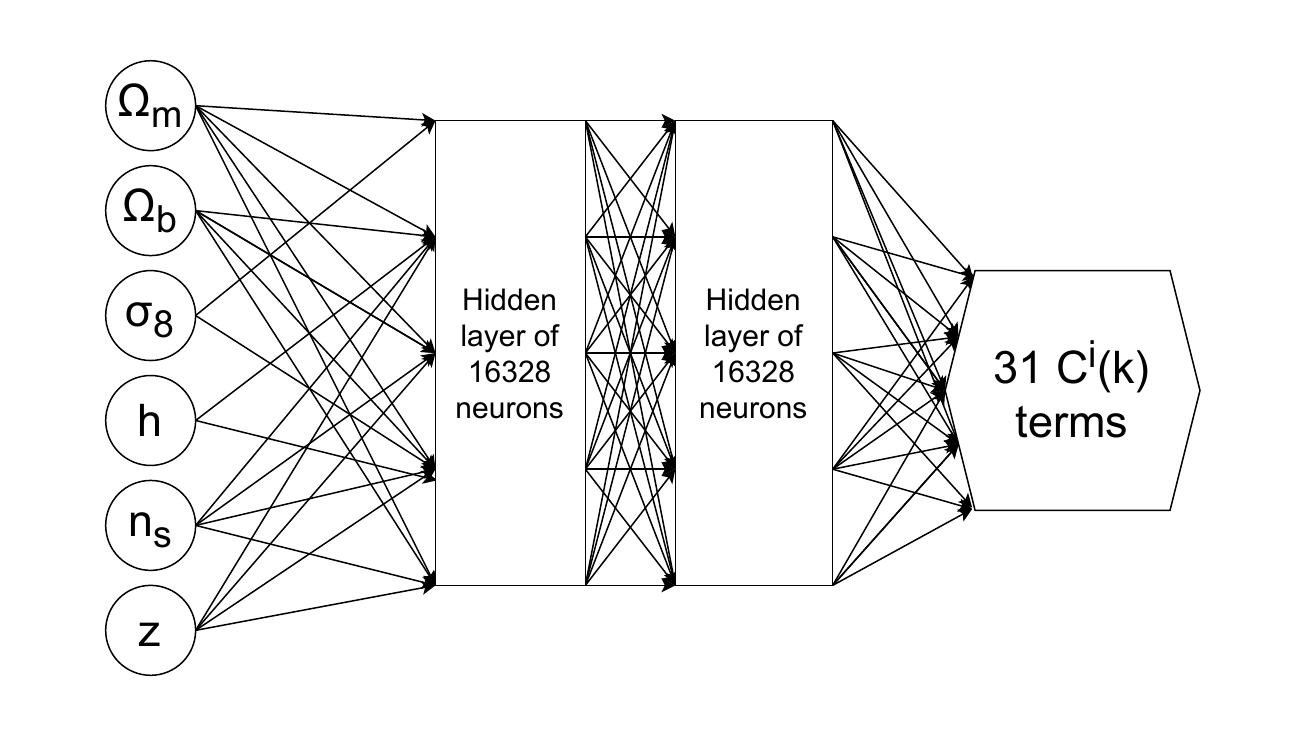}
    \caption{Architecture of the neural network emulator: The 6 $\Lambda$CDM cosmological parameters and the redshift $z$ are used as input parameters of a fully-connected neural network model composed of 2 hidden layers. The output of the neural network emulator are the predicted 31 bias-invariant terms that enter the PT predictions, binned in 50 bins of $k=[0.0,0.3]$.}
    \label{fig:nn-arch}
\end{figure}

Fig.~\ref{fig:nn-arch} presents the architecture of our neural network. The input parameters of the $\Lambda$CDM model are $\boldsymbol{\mathcal{C}} = \left\{\omega_{\rm cdm}, \omega_{\rm b}, \log\left[10^{10} A_s\right], n_s, h\right\}$ and the redshift $z$. For $w_o w_a$CDM, we just add the two parameters describing the dark energy equation of state, the present-day parameter $w_0$ and the time evolution parameter $w_a$. The training set 
comprises of 3000 (6000 for  $w_o w_a$CDM) samples drawn from a Latin hypercube\citep{Jin2005} featuring the six (eight for $w_o w_a$CDM) cosmological parameters and implemented using \texttt{SMT} toolkit\citep{saves2024smt}, and the redshift in the range $0 < z < 1.4$. All the datasets are generated using \texttt{MomentExpansion} module of \texttt{velocileptors}. Table~\ref{tab:priors} summarises the very broad flat un-informative priors used for the cosmological parameters when defining the Latin hypercube. 
We group the training data into a $31 \times 50$ matrix, where 31 is the number of bias-independent terms described in Section~\ref{sec:ME} and 50 is our fiducial choice for the
number of $k$ bins.

\begin{table*}
    \caption{Definitions and ranges of the parameters of the training set for the emulator.}
    \label{tab:priors}
    \centering
    \rowcolors{2}{white}{gray!15}
    \begin{tabular}{l c r}
        \hline
        Parameter & Interpretation & Prior range \\
        \hline
        $\omega_{\rm cdm}$ & Physical cold dark matter density parameter & [0.05, 0.30] \\
        $\omega_{\rm b}$ &  Physical baryon density parameter & [0.01, 0.04] \\
        $\log\left[10^{10} A_s\right]$ & The primordial normalization of the matter power spectrum at $k=0.05$  Mpc/h  & [2, 4]\\
        $n_s$ & Spectral index of the primordial power spectrum & [0.8, 1.1]\\
        $h$ & Normalized Hubble constant at $z=0$ & [0.5,0.8] \\
        $z$ & Redshift & [0.0, 1.4] \\
        $w_0$ & Present-day dark energy equation of state & [-0.5, -2] \\
        $w_a$ & Time evolution of the dark energy equation of state & [-3, 0.3] \\
        \hline
    \end{tabular}
\end{table*}


A feed-forward fully-connected model based on the machine learning framework \texttt{pytorch}\footnote{\url {https://github.com/pytorch/pytorch}} is created for each such matrix. We use 2 hidden layers of $16238$ neurons and the Gaussian Error Linear Units (GELU) activation function \citep{2016arXiv160608415H}, which can be represented as  
\begin{equation}
    \mathrm{GELU}(x) = 0.5 x \left[1+\text{erf}\left(\frac{x}{\sqrt{2}}\right) \right].
\end{equation}
The overall outputs (perturbation theory terms) and the inputs (cosmological parameters) $x_i$ are normalised $x_i\in[-1,1]$.

The training is done in batches of 128 for 5000 epochs, meaning that the training dataset is divided into groups of 128, where the elements of each group are then simultaneously passed through the neural network, and after that the weights are adjusted using backpropagation. These groups are called batches, and we do this until all of the possible groups have been used. That constitutes an epoch. This procedure is therefore repeated 5000 times.  The validation dataset consists of 1000 samples, constituting a hypercube with the same parameters as the training data. We minimise the L1 norm loss function defined by:
\begin{equation}
    \label{eq:loss}
    \mathcal{L} = \frac{1}{N} \sum_{i=0}^N |y^i_\mathrm{true} - y^i_\mathrm{predicted}|,
\end{equation}
with optimisation performed using \texttt{pytorch} realisation of the Adam optimiser \citep{2014arXiv1412.6980K}. The learning rate is set to $4\times10^{-7}$. We stopped the training after $~100$ epochs when the validation loss (which is the same as for the training) was not improving. The tests of the emulator were done separately, and are described in the following section.


\subsection{Performance of the emulator}

First, we assess the performance of the emulator in predicting the Legendre multipoles of the power spectrum defined as:
\begin{align}
    P_{\ell}(k) = \frac{(2\ell + 1)}{2} \int_{-1}^1 d\mu\ P(k,\mu)\mathcal{L}_{\ell}(\mu)
\end{align}
where ${L}_{\ell}(\mu)$ is the Legendre polynomial of order $\ell$. In this work, we consider the monopole $\ell=0$, the quadrupole $\ell=2$ and the hexadecapole $\ell=4$.

\begin{table}
   \caption{Ranges of the parameters used for the multipole testing.}
    \label{tab:priors_test}
    \centering
    \rowcolors{2}{white}{gray!15}
    \begin{tabular}{l l c r}
        \hline
        Parameter &  Range \\
        \hline
        $\omega_{\rm cdm}$ & [0.10, 0.14]  \\
        $\omega_{\rm b}$ & [0.01, 0.03]  \\
        $\log \left(10^{10}A_s\right)$ & [2.5, 3.5] \\
        $ h $ & [0.64, 0.72] \\
        $n_s$ & [0.9, 1.0]  \\
        $b_1$ & [-1, 3] \\
        $b_2$ & [-10, 10] \\
        $b_s$ & [-20, 20] \\
        $b_3$ & [-20, 20] \\
        \hline
    \end{tabular}
\end{table}

\begin{figure}
    \centering  \includegraphics[width=1.\linewidth]{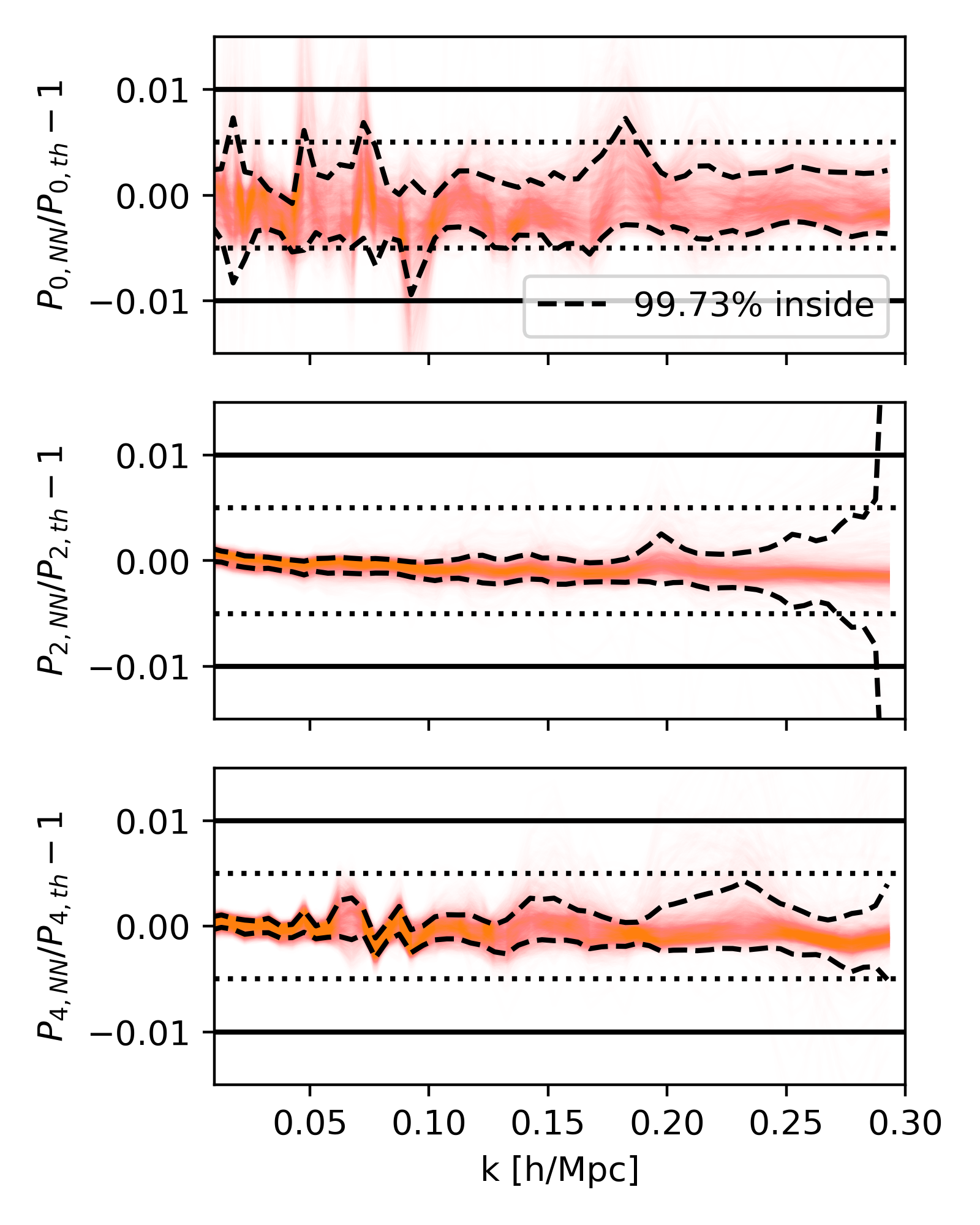}
    \caption{Comparison between the galaxy redshift space power spectrum multipoles of the emulator $P_{\ell,\textrm{NN}}$ and of the theoretical version $P_{\ell,\textrm{th}}$ for $\ell=0$ (top), $\ell=2$ (middle) and $\ell=4$ (bottom), for $\Lambda$CDM. The dashed curves represent the 3$\sigma$ scatter and the red curves the individual realisations.}
    \label{fig:perf-multipoles}
\end{figure}

In order to assess the performance of the emulator at the level of the multipoles, we generate $N=10\,000$ sets of the cosmological and nuisance parameters taken from the ranges given in Table \ref{tab:priors_test}. Then, we produce the multipoles using both the original \texttt{velocileptors} code and our emulator.
Fig.~\ref{fig:perf-multipoles} shows the ratio of the neural network LPT emulator multipole $P_{l,\textrm{NN}}$ to the theoretical prediction from \texttt{velocileptors} $P_{l,\textrm{th}}$ for the monopole (top), quadrupole (middle) and hexadecapole (bottom), for $\Lambda$CDM. The dashed curves show the 3$\sigma$ scatter. Up to $\kmax=0.25\hMpc$, the overall multipoles computed from the emulator agree with the ones from the reference analytic version at below 0.5\% at 3$\sigma$, which means below 0.2\% at 1$\sigma$.

\begin{figure}
    \centering  \includegraphics[width=1.\linewidth]{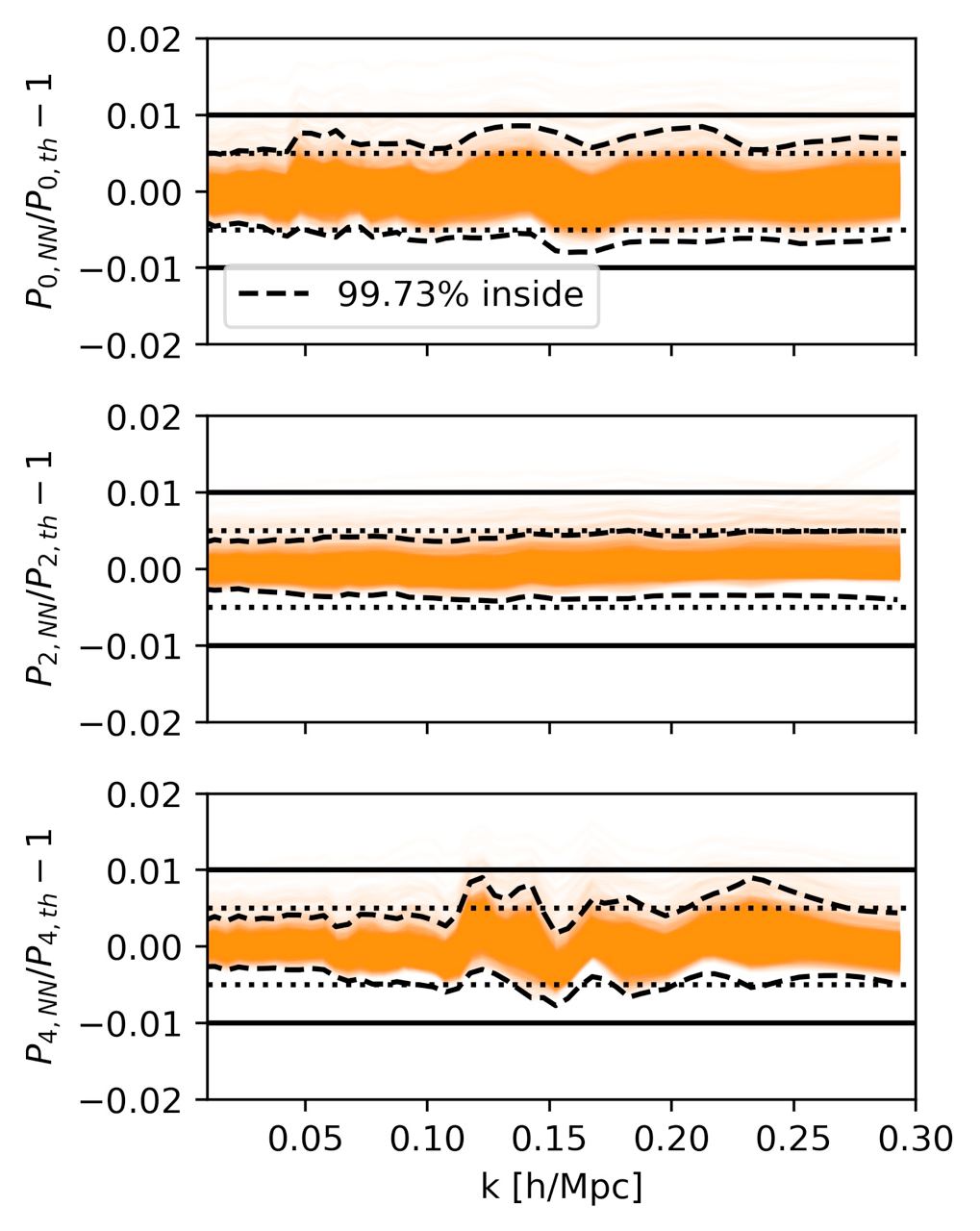}
    \caption{Same as for Fig.~\ref{fig:perf-multipoles} but for $w_o w_a$CDM. We recover a performance somewhat worse than that for the $\Lambda$CDM case, due to the 2 additional parameters and despite additional samples in the training set, but with $\sim0.5\%$ in precision at $3\sigma$}.
    \label{fig:perf-multipoles-wowa}
\end{figure}

Fig.~\ref{fig:perf-multipoles-wowa} shows the same information as Fig.~\ref{fig:perf-multipoles} but for $w_o w_a$CDM model. We recover a similar accuracy even for this extended cosmological model with the emulator predicting the multipoles at a precision below 0.2\% at 1$\sigma$ up to $\kmax=0.25\hMpc$.

\begin{figure}
    \centering
    \includegraphics[width=1.\linewidth]{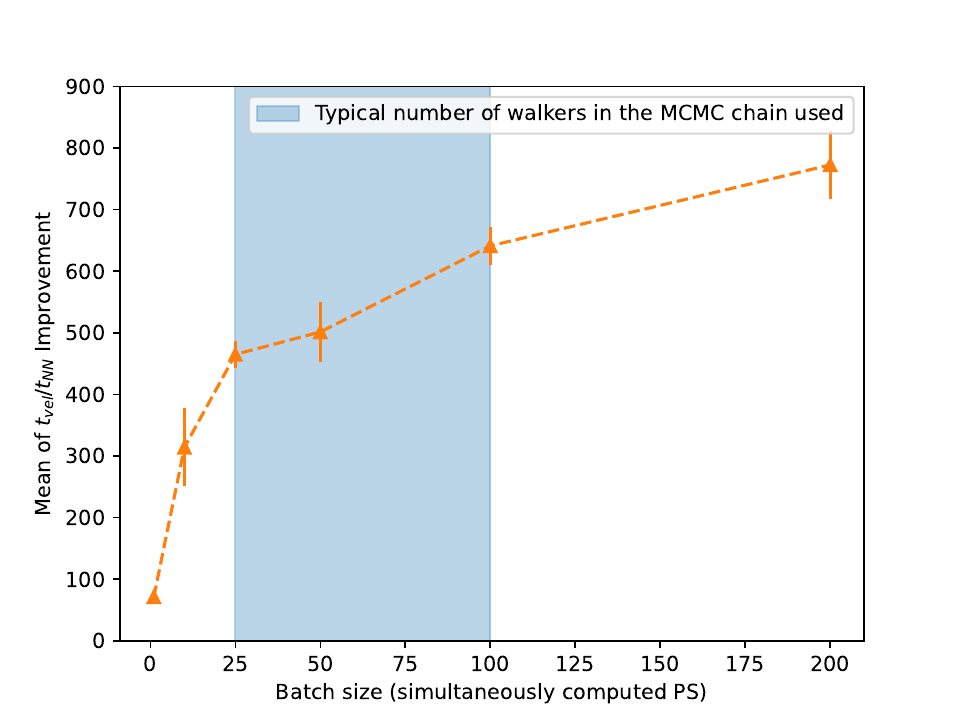}
    \caption{Speed performance of the neural network emulator with respect to the original code as a function of the number of simultaneously computed multipoles. The ratio of computation time  for the time with original code to that of our emulator is plotted against the batch size: number of simultaneously computed non-linear power spectra.}
    \label{fig:perf-time}
\end{figure}

We test the improvement in speed to compute the power spectrum multipoles by generating 50 batches of multipoles with the variable number of multipoles in each $n_b = [0,10,25,50,100,200]$, such that we can estimate the performance boost as a ratio between the elapsed time for their production with the original code to the time taken by the emulator. The corresponding proportion is then plotted against the number of multipoles in a single batch in Fig.~\ref{fig:perf-time}. We attribute most of the speed growth with increasing the batch size as due to the parallelisation over Graphics Processing Units (GPU), a feature that the original \texttt{velocileptors} software does not support due to the very sequential nature of its code. 


\section{Cosmological inference}
\label{sec:inference}

In this section, first we describe the simulations we use to compare the cosmological constraints from the emulator described in the previous section and its analytic version. Then, we present the results of the cosmological inference when using either the emulator or the original EFT code.

\subsection{DESI-like simulations}
We use the AbacusSummit suite of cosmological N-body simulations \citep{2021MNRAS.508.4017M} that were run with the Abacus N-body code \citep{2021MNRAS.508..575G}. We use the 25 base simulations each with 330 billion particles in a 2~$(h^{-1}\text{Gpc})^3$ volume which corresponds to a mass resolution of $2\times10^9 M_{\odot}/h.$ The baseline cosmology is Planck 2018~\citep{Planck2020}, specifically the mean of \texttt{base\_plikHM\_TTTEEE\_lowl\_lowE\_lensing}. We consider two snapshots $z=0.5$ and $z=0.8$ and only cubic boxes.

In order to test the performance of the emulator in extracting robust cosmological inference, we create 3 sets of Abacus galaxy mocks: two Luminous Red Galaxies (LRG) boxes at $z=0.5$, at $z = 0.8$ and one Emission line Galaxies (ELG) box at $z = 0.8$, which allows us to test both the effect of redshift evolution and the dependence on the nature of the tracer. We populate the dark matter halos with galaxies with similar clustering properties to those found by DESI, using the Halo Occupancy Distribution formalism which connects the probability for a halo of a given mass to host a galaxy. In this formalism, we treat separately the central galaxies located at the centre of the halo and satellite galaxies.
We follow the prescriptions and HOD parameter values that are tuned on DESI Early Data Release \citep[DESI EDR,][]{2023arXiv230606308D}. More precisely, for LRG we use the results from \cite{2023arXiv230606314Y} and for ELG the ones from \cite{2023JCAP...10..016R}.

The probability for a halo of mass $M_h$ to host a central LRG galaxy is given by:
\begin{equation}
    P_{\mathrm{cen, LRG}}(M_{h}) = \frac{1}{2} p_{\mathrm{max}}\,\mathrm{erfc}\left(\frac{\log_{10}(M_{\mathrm{cut}}/M_h)}{\sqrt{2}\sigma_M} \right)
\end{equation}
where $M_h$ is the halo mass, $\log_{10}M_{\mathrm{cut}}$ corresponds to the mass where only half of the halos hosts a central galaxy, $\log \sigma_M$ controls the width of the transition from hosting zero to one central galaxy and $p_{\mathrm{max}}$ controls the saturation level of the occupation probability or in other words it can be seen as the maximal probability that a halo hosts a central galaxy.

The model of central galaxies HOD is more complex for ELGs. We are using the one called the High Mass Quenched model as proposed in \cite{Alam2020_HOD}, based on a skewed distribution, allowing for a reduction of the central galaxies in higher mass halos. It introduces parameters $Q$, setting the quenching efficiency for higher mass halos and $\gamma$ controlling the skewness of the distribution. The overall HOD model for the ELG central galaxies can be written as:
\begin{equation}
    P_{\mathrm{cen,ELG}}(M_{h}) = 2A \phi(M_h)\Phi(\gamma M_h) + \frac{1}{2Q}\left[1+\mathrm{erf}\left(\frac{\log_{10}(M_{h}/M_{\mathrm{cut}}}{0.01} \right)\right]
\end{equation}
where:
\begin{equation}
    \phi(x) = \mathcal{N}(\log_{10}M_{\mathrm{cut}},\sigma_M)
\end{equation}
\begin{equation}
    \Phi(x) = \int^x_{-\infty}\phi(t) dt = \frac{1}{2}\left[1+\mathrm{erf}\left(\frac{x}{\sqrt{2}}\right)\right]
\end{equation}
\begin{equation}
    A = \frac{p_{\mathrm{max}} - 1/Q}{\max\left(2\phi(x)\Phi(\gamma x)\right)}.
\end{equation}

The number of the satellite galaxies $N_\textrm{sat}$ for both LRG and ELG is given by:
\begin{equation}
    \langle N_{\rm sat} \rangle (M_{h}) = \left(\frac{M_h - \kappa M_{\rm cut}}{M_1}\right)^\alpha
\end{equation}
where $M_1$ characterizes a typical mass of the halo hosting 1 satellite galaxy, and $\kappa M_\textrm{cut}$ controls the minimal mass for a halo to host a satellite galaxy.
In order to create the ELG mocks we use the modified model taken from \cite{2023JCAP...10..016R}, where $Q$ tends to infinity.
We use the public code package \texttt{ABACUSHOD}~\citep{2022MNRAS.510.3301Y} which is part of the \texttt{ABACUSUTILS} package~\footnote{\url{https://github.com/abacusorg/abacusutils}} to apply these HOD prescriptions to the dark matter halos of the AbacusSummit simulations and the values of the HOD parameters for LRG and ELG are summarised in Table~\ref{tab:HOD}.

\begin{table*}
    \caption{Definitions and ranges of the galaxy-halo connection parameters for the the simulations used to test the emulator.}
    \label{tab:HOD}
    \centering
    \rowcolors{2}{white}{gray!15}
    \begin{tabular}{l l c r r}
        \hline
        Parameter & Interpretation & LRG $z=0.5$ & LRG $z=0.8$ & ELG $z=0.8$ \\
        \hline
        $\log(M_{\rm cut})$ & Minimum halo mass to host a central & 12.79 & 12.64 & 11.75 \\
        $\log(M_1)$ & Typical halo mass to host one satellite & 13.88  & 13.71 & 19.83 \\
        $\sigma_M$ & Scatter around the mean halo mass or $M_{\rm cut}$? & 0.21  & 0.09 & 0.31 \\
        $\alpha$ & Power-law index for the mass dependence of the number of satellites& 1.07 & 1.18 & 0.72\\
        $\kappa$ & Parameter that modulates the minimum halo mass to host a satellite & 1.4 & 0.6 & 1.8\\
        $p_{\rm max}$ & Maximal probability of a galaxy to occupy a halo& 1 & 1 & 0.08 \\
        $\gamma$ & Quenching efficiency & n/a & n/a & 1.39\\
        $\alpha_c$ & Central velocity bias & 0.33 & 0.19 & 0.19\\
        $\alpha_s$ & Satellite velocity bias & 0.80 & 0.95 & 1.49\\
        \hline
    \end{tabular}
\end{table*}

\subsection{Methodology}
For each of the 25 Abacus mocks, we compute the 2-point redshift-space power spectrum multipoles ($\ell=0,2,4$) and the associated window function for the cubic boxes using \texttt{pypower}\footnote{\url{https://github.com/cosmodesi/pypower}}. The code is based on the methodology described in \cite{Yamamoto} and \cite{Hand2017}. The density $\delta(x)$ is computed on a mesh of size $512^3$, then using FFT, first we can obtain the quantities $\delta_{\ell}$ defined as:
\begin{equation}
    \delta_{\ell} (k) =\int \frac{d^3x}{(2\pi)^3} e^{-i\boldsymbol{k}\cdot\boldsymbol{x}} \delta_s(\boldsymbol{x})\mathcal{L}_{\ell}(\hat{k}\cdot\hat{x})
\end{equation}

Those terms later can be combined into the power spectrum multipoles as:
\begin{equation}
    P_{\ell}(k) = (2\ell+1)\int \frac{d\Omega}{4\pi}\delta_{\ell}(k)\delta_0(-k)
\end{equation}
where $\delta_0=\delta_{l=0}$ and $\Omega$ is the solid angle. As we are planning to apply it to boxes, we assume the sky to be flat, and take the LOS to be along a box side.  


Once we measure the multipoles, we use the mean of 25 simulations to create a Gaussian analytic covariance matrix following the methodology in \cite{Grieb2015}:
\begin{equation}
\begin{split}
    C^{\mathcal{G}}_{\ell_1,\ell_2}(k_i,k_j) = \frac{2(2\pi)^4(2\ell_1+1)(2\ell_2+1)}{V_s V_{k}^2}\delta_{ij}\times \\ \times\int^{k_i+\Delta k}_{k_i-\Delta k} k^2  \left[ \int ^{1}_{-1}\left( P(k,\mu)+\frac{1}{\overline{n}}\right)\mathcal{L}_{\ell_1}(\mu)\mathcal{L}_{\ell_2}(\mu) d\mu\right] dk
    \end{split}
\end{equation}

where $V_s$ is the volume of a sample, $V_k$ is the volume of a shell in k-space, and the $\frac{1}{\overline{n}}$ term represents the shot-noise contribution, which we assume to be negligible.

Once we have both the multipoles and the covariance, we can compute the log-likelihood $L(p_1, ..., p_n)$ with respect to the chosen theory and a set of cosmological and nuisance parameters as:
\begin{equation}
\begin{split}
   \log(L(p_1,...,p_n)) = \sum_{\ell_{1} \ell_{2}} \sum_{i,j} \left(P_{\ell_1}(k_i) - P^{th}_{\ell_1}(k_i)\right) \times \\ \times\left(\Sigma^{-1}\right)^{\ell_1 \ell_2}_{i,j} \left(P_{\ell_2}(k_j) - P^{th}_{\ell_2}(k_j)\right)
\end{split}
\end{equation}
where $P^{th}_{\ell}$ are the theoretical predictions of the power spectrum multipoles, $\{p_1, ..., p_2\}$ are the parameters of the model including the cosmological parameters, but also the galaxy bias and nuisance terms. We use \texttt{emcee}\footnote{\url{https://emcee.readthedocs.io/en/stable/}} package \citep{2013PASP..125..306F} for the MCMC sampling in order to infer the cosmological parameters of interest. All of the MCMC chains are run just to the convergence, which is tracked using integrated auto-correlation time for which we require to be $\tau > N/100$ \citep{Sokal}, where $N$ is the number of steps that walkers in the sampler have made.

\subsection{Consistency test of the multipoles}

\begin{figure}
    \centering
    \includegraphics[width=1.\linewidth]{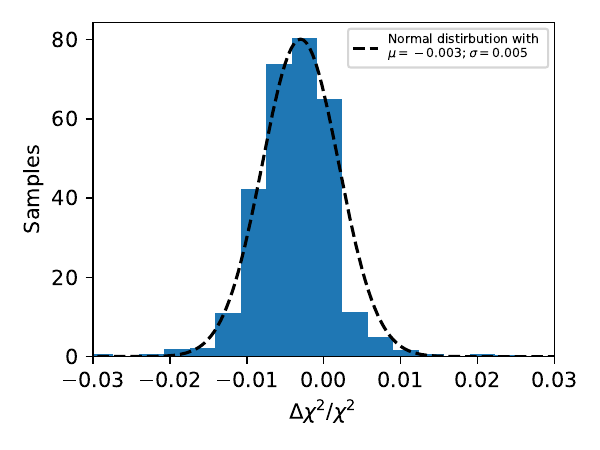}
    \caption{Distribution histogram of the ratios of the difference in $\chi^2$ obtained with our emulator and the original code to the $\chi^2$ obtained with the original code.}
    \label{fig:perf-chi2}
\end{figure}

\begin{figure}
    \centering
    \includegraphics[width=1.\linewidth]{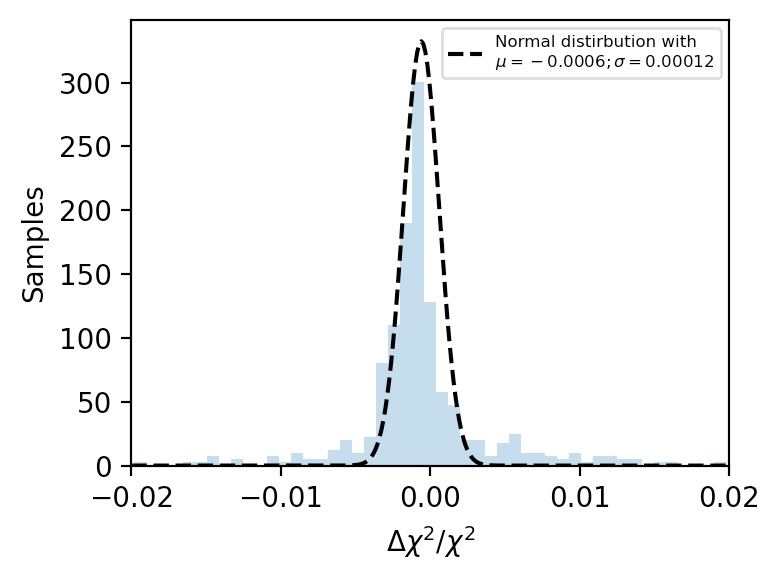}
    \caption{Same as for Fig.~\ref{fig:perf-chi2} but for $w_o w_a$CDM.}
    \label{fig:perf-chi2-wowa}
\end{figure}

We perform a consistency test at the level of the multipoles between the measured and predicted multipoles using the same test set as the one used to produce Fig.~\ref{fig:perf-multipoles} for $\Lambda$CDM and Fig.~\ref{fig:perf-multipoles-wowa} for $w_o w_a$CDM. To do so, we define $\Delta \chi^2 = \chi^2_{\mathrm{emu}}-\chi^2_\textrm{th}$ where $\chi^2$ can be defined from a Gaussian likelihood as 
\begin{equation}
    \chi^2 = - 2 \log\left(L(p_1,...,p_n)\right) + C
\end{equation}
where $C$ is a normalisation constant for the likelihood distribution. This test allows us to estimate the difference in terms of likelihoods generated using \texttt{velocileptors} and our emulator by inputting the same cosmological and bias parameers into both models, with the parameters being sampled from uniform distribution defined in Table \ref{tab:priors_test}. For the `data' vector we use the mean of 25 LRG $z=0.8$ mocks, and for the covariance we use the analytic covariance for the same sample. We then compute the $\chi^2$ for each model, and then take their difference $\Delta\chi^2$ and divide it by the value obtained from the \texttt{velocileptors} model.
Fig.~\ref{fig:perf-chi2} and Fig.~\ref{fig:perf-chi2-wowa} show the distribution of $\Delta \chi^2 / \chi^2$ for this test within $\Lambda$CDM and $w_o w_a$CDM respectively. The standard deviation of this statistic is $\sigma(\Delta \chi^2 / \chi^2)= 0.005$ with a mean of $\mu(\Delta \chi^2 / \chi^2)= 0.003$. This shows that for the vast majority of cases we will obtain the correct likelihood with a sub-percent level of precision.

\subsection{Cosmological inference: comparison between the emulator and \texttt{velocileptors}}

\begin{figure*}
    \captionsetup[subfigure]{labelformat=empty}
    \begin{subfigure}{.5\textwidth}
    \centering
    \includegraphics[height=8cm]{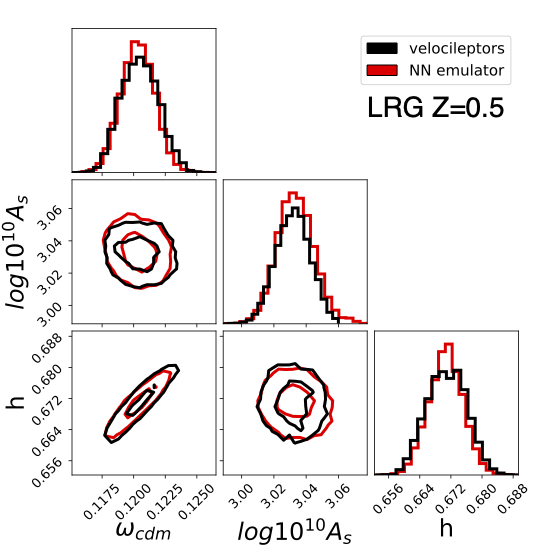}
    \end{subfigure}%
    \begin{subfigure}{.5\textwidth}
    \centering
    \includegraphics[height=8cm]{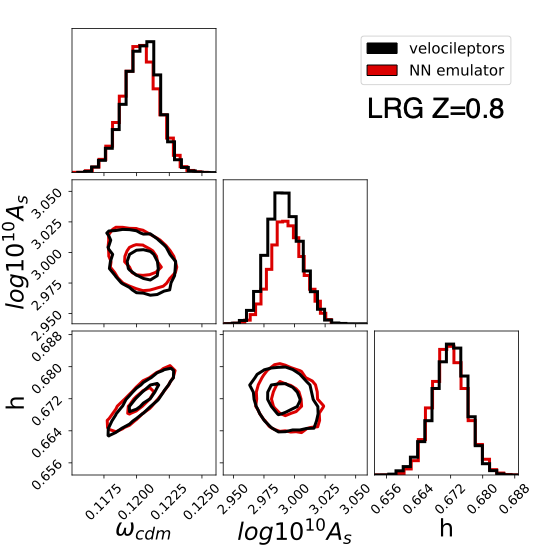}
    \end{subfigure}
    \begin{subfigure}{.5\textwidth}
    \centering
    \includegraphics[height=8cm]{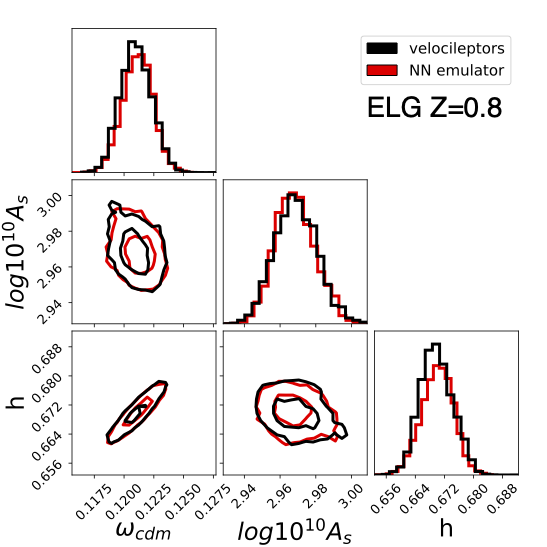}
    \end{subfigure}%
    \caption{Comparison of cosmological constraints obtained with the neural network emulator (red) and with \texttt{MomentExpannsion} (black) when fitting the mean of the 25 Abacus mocks for the three configurations.}
    \label{fig:mcmc}
\end{figure*}

We assess the performance of the neural network LPT emulator in inferring the cosmological parameters with respect to the original LPT code by fitting the mean of the 25 Abacus mocks for the three sets: DESI LRG-like at $z=0.5$, DESI LRG-like at $z=0.8$ and DESI ELG-like at $z=0.8$. The fits are performed for $\ell=0,2,4$ multipoles of the power spectrum in k-ranges of $[0.02,0.2]$ in bins of $\Delta k = 0.005$.Fig.~\ref{fig:mcmc} shows the cosmological constraints obtained from full modelling using either the neural network LPT emulator (red) or \texttt{MomentExpansion} module of \texttt{velocileptors} (green). The dashed curves represent the 1$\sigma$ error from each model. Both models yield very consistent results, both for the best-fit values and the uncertainty.



\begin{figure}
    \centering
    \includegraphics[width=0.8\linewidth]{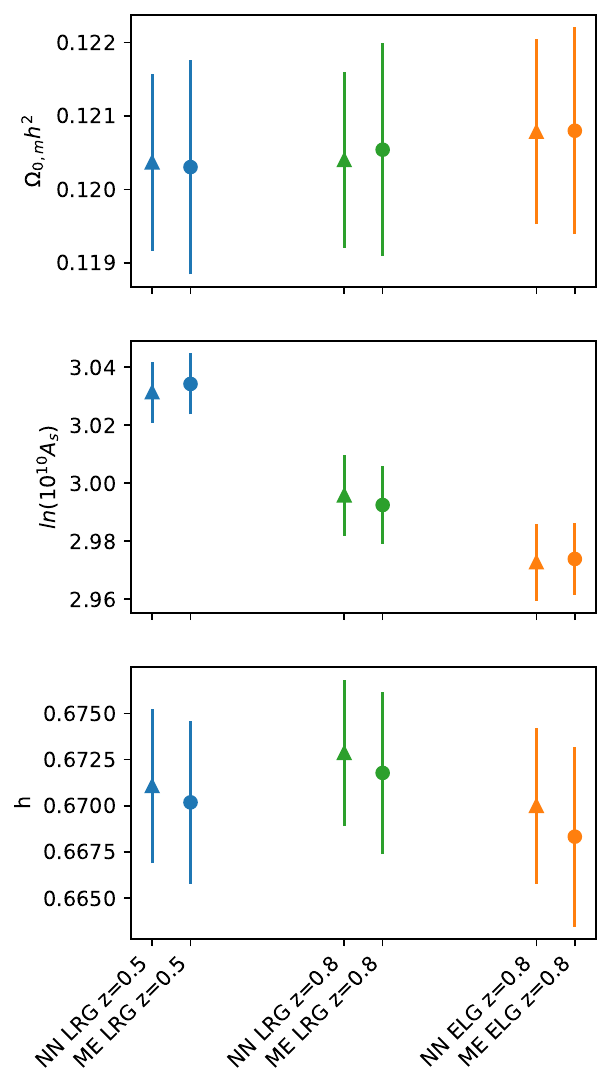}
    \caption{The cosmological parameters obtained from Full-Modelling fits with the neural network emulator and with the original code obtained from the mean of different mock types with rescaled covariance matrix.}
    \label{fig:cosmo-summary}
\end{figure}

We summarize the results of the comparison between the neural network emulator (triangles) and the analytic code (circles) in Fig.~\ref{fig:cosmo-summary} where we show the best-fit values and 1$\sigma$ error for the three main cosmological parameters that are well constrained with Full-Shape analysis. We can see that both methods yield very consistent and similar results with less than 0.3$\sigma$ for the largest difference seen on $h$.

\subsection{Cosmological inference: comparison between the emulator and the truth}

\begin{figure}
    \centering
    \includegraphics[width=1.\linewidth]{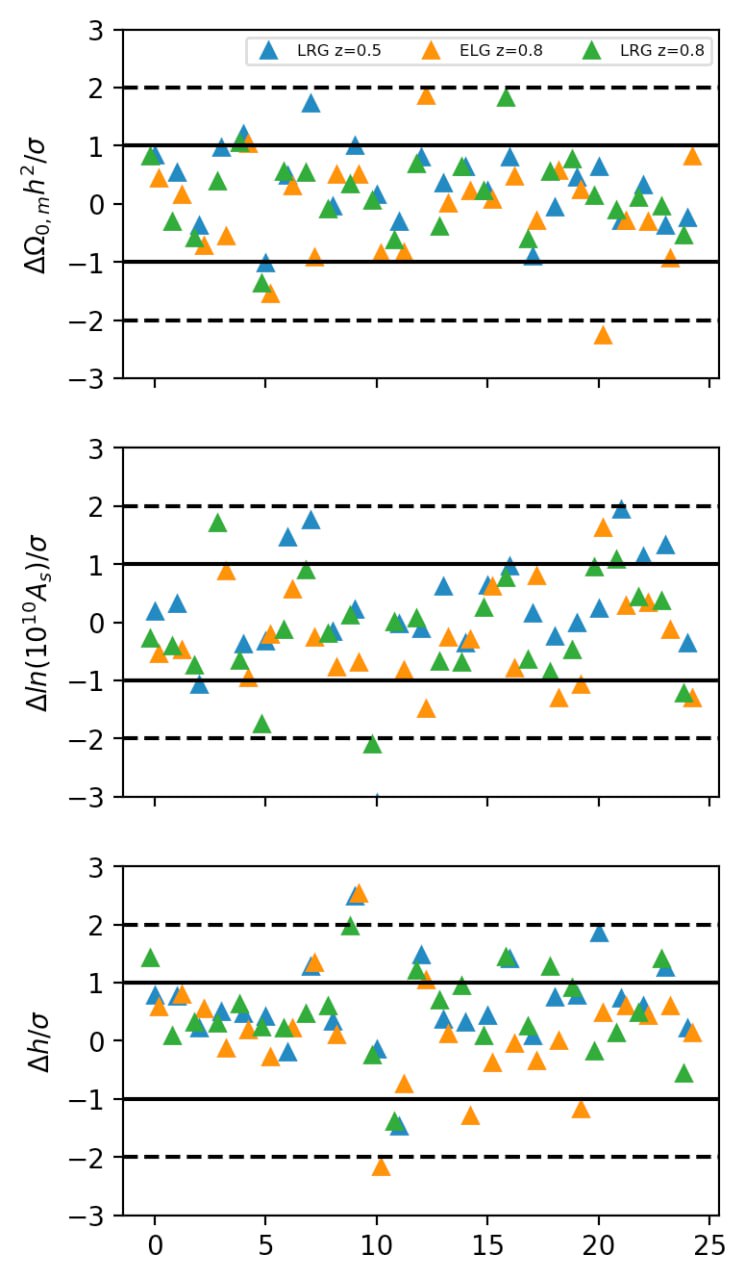}
    \caption{The deviation of different cosmological parameters in terms of the error $\sigma$ from the expected theoretical value obtained from the individual mock fits for three mock types: blue for LRGs with $z=0.5$, green for LRGs with $z=0.8$ and orange for ELGs with $z=0.8$.}
    \label{fig:lrgz0.8-ind}
\end{figure}

We also want to assess the performance of the emulator with respect to the expected values given the cosmological model that was used for the simulations. In order to do that, we fit the 25 individual realisations for each case and in Fig.~\ref{fig:lrgz0.8-ind}, we show the results for the cosmological parameters by plotting the difference between measured and truth divided by the error on the measured parameter as a function of mock number. Blue triangles represent the results for LRG at $z=0.5$, green ones for LRG $z=0.8$ and orange ones for ELG at $z=0.8$. 
All the results are consistent with the expected truth values within 1-2 $\sigma$, which further validates the ability of the emulator to recover precise and unbiased cosmological constraints.

\section{Conclusions}
\label{sec:concl}

We have developed a neural network-based emulator for the non-linear redshift space galaxy power spectrum that is tailored for galaxy clustering analysis that relies on the Effective Field Theory of Large-Scale Structures (EFTofLSS). It takes as input the cosmological parameters {} and the redshfit and emulates only the cosmology-dependent terms of a 1-loop Lagrangian perturbation theory which, in this work, is taken to be the one implemented in the public code \texttt{velocileptors} \citep{velocileptors1,velocileptors2}. The other terms of the theory (bias expansion, effective field theory terms also called counter-terms and stochastic terms) are kept analytical and combined with the output of the neural network in order to obtain the non-linear redshift space galaxy power spectrum. First and foremost, it bypasses the need to generate a linear matter power spectrum from a Boltzmann code such as \texttt{CLASS} or \texttt{CAMB} which is the most computationally demanding task. As a consequence, it enables acceleration of the full inference pipeline by a factor of $\mathcal{O}(10^3$). Moreover, there is an additional speed-up that comes when running the emulator on GPU which is not possible with the original \texttt{velocileptors} code. We have also shown that the accuracy of the emulator meets the requirements for the new generation of galaxy surveys such as DESI \citep{2016arXiv161100036D}. First, we have checked the residuals between the emulated and analytical multipoles of the redshift space galaxy power spectrum, then we have examined the$\chi^2$ difference i of the likelihood evaluation between the emulated and predicted multipoles. Eventually, we have performed a full-inference analysis in order to compare the posterior on the cosmological parameters between the two methods. For the last two tests, we have created three sets of mocks that mimic DESI galaxy samples using the public \textsc{AbacusSummit} N-body simulations: LRG at $z=0.5$, LRG at $z=0.8$ and ELG at $z=0.8$ 

In addition to the significant speed-up that the emulator provides with respect to the original analytical code, the emulator has also other key advantages:
\begin{itemize}
    \item {\bf accurate}: we have showed that our emulator can predict the multipoles of the redshift space galaxy power spectrum with 0.5\% accuracy at $k_\mathrm{max}$=0.25$\hMpc$ at 3$\sigma$ within $\Lambda$CDM. 
    \item {\bf flexible}: We have found similar performance of the emulator independently of the redshift and of the nature of the tracer. We tested the case of DESI-like LRG and ELG, but we stress that any galaxy sample could be used. Because we decided to emulate only the bias-independent terms that depend on the cosmological model and combine a posteriori with the bias expansion terms and nuisance terms, the emulator is very flexible with respect to any type of galaxy sample considered. Moreover, the inclusion of the redshift in the training implies that no additional re-training is needed from the user's point of view. Therefore,, although we provide all the tools necessary to train the emulator, this is not needed if the emulator is used in the cosmological and redshift ranges indicated in Table~\ref{tab:priors}.
    \item {\bf differentiable}: by construction the emulator is fully differentiable which makes it useful for gradient-based inference such as Hamiltonian Monte Carlo \citep[HMC, e.g.][]{2017arXiv170102434B} which is more efficient for high dimensional parameter space sampling.
    \item {\bf beyond-$\Lambda$CDM}: we have also considered an extension of the cosmological parameter space by including the time-parametrisation of the dark energy equation of state $w_o$,$w_a$. We obtain a slightly worse performance as in the $\Lambda$CDM case with the emulator reproducing the multipoles with around 0.5\% precision at 3$\sigma$ up to $k_\mathrm{max}$=0.25$\hMpc$ while using the same architecture with additional samples in the training set.
\end{itemize}

The use of neural networks for cosmological power spectra emulation has been more and more common. \texttt{CosmoNET} was one of the NN-based emulator developed for accelerating the calculation of CMB power spectra, matter transfer functions and likelihoods \citep{2008MNRAS.387.1575A}. More recently, \cite{2022MNRAS.511.1771S} developed \texttt{COSMOPOWER} that emulates both the CMB power spectra and matter power spectrum computed by Boltzmann codes. We found similar gain in speed and performance for the LSS part with the main difference being that in our case, we obtain directly a prediction of the non-linear redshift space galaxy power spectrum by emulating the bias-invariant terms of the LPT kernels and combining them with a bias expansion model and nuisance terms including counterterms from EFT.
Therefore, our $k$- and $z$-ranges are limited to the validity of EFT/LPT in the quasi-linear regime and to the typical redshift range probed by galaxy clustering ($0 < z < 1.5$). In \cite{2022JCAP...04..056D}, they also use neural networks as fast surrogate for the non-linear redshift space galaxy power spectrum, but also for the real space galaxy, galaxy-matter and matter-matter power spectra so that it can be used for both galaxy clustering and weak lensing analyses. However, their cosmological parameter space is more restricted than ours with $n_s$ fixed and without including the redshift $z$ in the training set, which means the user has to re-train the emulator for each redshift considered. Moreover, they include galaxy bias terms, counterterms and stochastic terms in the training set. Although their prior ranges for these non-cosmological parameters are broad, their emulator cannot be used as it is for more exotic bias expansion models which would require a re-training operation. 

The development of machine learning algorithms as surrogate models for cosmological observables has been motivated by the need to decrease the computational cost of parameter estimation. This is more and more true as the models increase in complexity with additional nuisance parameters in order to meet the stringent accuracy requirements imposed by more precise cosmological measurements. In this work, we have focused on speeding up the prediction of the non-linear redshift space galaxy power spectrum which implies a speed-up of the full inference pipeline from direct fitting of the galaxy two-point statistics. In a future work, we will use this neural network emulator to perform a multi-tracer analysis of the DESI BGS DR1 (Trusov et al. in prep). Accelerating the inference pipeline as proposed in this work becomes even more crucial for multi-tracer as, in the case of the DESI BGS, we analyse jointly blue, red and cross power spectra or correlation functions, which increases significantly the dimensionality of the parameter space and the computational expense of parameter estimation. 


\section*{Acknowledgements}

ST and PZ thank Peder Norberg for useful discussions about this work.
ST and PZ acknowledge the Fondation CFM pour la Recherche for their financial support. 
PN and SC acknowledge STFC funding ST/T000244/1 and ST/X001075/1.

This work used the DiRAC@Durham facility managed by the Institute for Computational Cosmology on behalf of the STFC DiRAC HPC Facility (www.dirac.ac.uk). The equipment was funded by BEIS capital funding via STFC capital grants ST/K00042X/1, ST/P002293/1, ST/R002371/1 and ST/S002502/1, Durham University and STFC operations grant ST/R000832/1. DiRAC is part of the National e-Infrastructure.

This material is based upon work supported by the U.S. Department of Energy, Office of Science, Office of High Energy Physics of U.S. Department of Energy under grant Contract Number DE-SC0012567, grant DE-SC0013718, and under DE-AC02-76SF00515 to SLAC National Accelerator Laboratory, and by the Kavli Institute for Particle Astrophysics and Cosmology. The computations in this paper were run on the FASRC Cannon cluster supported by the FAS Division of Science Research Computing Group at Harvard University, and on the Narval cluster provided by Compute Ontario (computeontario.ca) and the Digital Research Alliance of Canada (alliancecan.ca). In addition, this work used resources of the National Energy Research Scientific Computing Center (NERSC), a U.S. Department of Energy Office of Science User Facility located at Lawrence Berkeley National Laboratory, operated under Contract No. DE-AC02-05CH11231.

The \textsc{AbacusSummit} simulations were run at the Oak Ridge Leadership Computing Facility, which is a DOE Office of Science User Facility supported under Contract DE-AC05-00OR22725.

\section*{Data Availability}
The data underlying this article are available in \url{https://abacusnbody.org}.

The emulator codes can be accessed using the following github repository: \url{https://github.com/theonefromnowhere/ME_NN_Emu}.



\bibliographystyle{mnras}
\bibliography{biblio}







\bsp	
\label{lastpage}
\end{document}